# Natural layered phlogopite dielectric for ultrathin two-dimensional optoelectronics

*Thomas Pucher[1*], Julia Hernandez-Ruiz[1], Guillermo Tajuelo-Castilla[2], José Ángel Martín-Gago[2], Carmen Munuera[1] and Andres Castellanos-Gomez[1*]*

[1]*2D Foundry Research Group. Instituto de Ciencia de Materiales de Madrid (ICMM-CSIC), Madrid, E-28049, Spain.*

[2]*ESISNA Research Group. Instituto de Ciencia de Materiales de Madrid (ICMM-CSIC), Madrid, E-28049, Spain.*

thomas.pucher@csic.es

andres.castellanos@csic.es

ABSTRACT

The integration of high-dielectric-constant (high-κ) materials with two-dimensional (2D) semiconductors is promising to overcome performance limitations and reach their full theoretical potential. Here we show that naturally occurring phlogopite mica, exfoliated into ultrathin flakes, can serve as a robust high-κ dielectric layer for transition metal dichalcogenide-based 2D electronics and optoelectronics. Phlogopite's wide bandgap (~4.8 eV), high dielectric constant (~11), and large breakdown field (>10 MV cm$^{-1}$) enable transistors with subthreshold swings down to 100 mV dec$^{-1}$, minimal hysteresis (~30–60 mV) and interface trap densities comparable to state-of-the-art oxide dielectrics. Moreover, phototransistors built upon monolayer molybdenum disulfide (MoS$_2$) and phlogopite exhibit responsivities up to 3.3×10$^4$ AW$^{-1}$ and detectivities near 10$^{10}$ Jones, surpassing devices based on conventional gate insulators. We further demonstrate the versatility of this natural dielectric by integrating phlogopite/MoS$_2$ heterostructures into NMOS inverters, showcasing robust voltage gains and low-voltage operation. Our findings establish phlogopite as a promising, earth-abundant dielectric for next-generation 2D transistor technologies and high-performance photodetection.

**INTRODUCTION**

Two-dimensional (2D) materials have attracted considerable attention in electronics and optoelectronics due to their atomic thinness, and outstanding properties such as high carrier mobility and tunable bandgaps[1,2]. Among these, transition metal dichalcogenides (TMDs), particularly $MoS_2$, have shown remarkable potential in ultrathin transistor and photodetector applications[3–10]. Nonetheless, achieving optimal performance in 2D electronic devices needs integration with suitable gate dielectrics characterized by high dielectric constants (κ), low leakage currents, large dielectric breakdown strength, and minimal interfacial defects[11].

Traditionally employed high-κ oxides like hafnium oxide ($HfO_2$) or aluminium oxide ($Al_2O_3$) suffer from inherent limitations such as amorphous structure-related defects, surface roughness, and deposition-induced damage to the sensitive surfaces of TMDs. Such drawbacks significantly degrade carrier mobility, introduce hysteresis, and diminish the subthreshold swing, critically restricting device efficiency and reliability. Recently, crystalline dielectrics, such as $SrTiO_3$[12,13], $BaTiO_3$[14], $Bi_2SeO_5$[15,16], $Sb_2O_3$[17,18] or $MgNb_2O_6$[19], have been explored to overcome these challenges. Despite their advanced electrical performance improvements, these works fail to acknowledge the optoelectronic promises of 2D semiconductors. Perovskite oxides, such as $Sr_2Nb_3O_{10}$, have been proposed to achieve dual functionalities of dielectric gating and photodetection[20]. However, these materials present challenges in ease of fabrication/assembly and limited achievable responsivities compared to other heterostructures.

While naturally occurring layered dielectrics, such as micas, have been proposed to overcome such fabrication complexities and achieve high performing optoelectronic 2D devices[21–24], so far, such results have not been reported.

Here, we introduce naturally occurring layered phlogopite mica as a highly promising alternative dielectric that addresses many of these limitations. Phlogopite mica, mechanically exfoliated from bulk crystals, exhibits exceptional dielectric characteristics, including a high dielectric constant (~11), robust dielectric breakdown fields (>10 MVcm$^{-1}$), and a wide optical bandgap (4.8 eV). Our devices, integrating phlogopite with MoS$_2$, demonstrate outstanding electrical performance with significantly reduced hysteresis, steep subthreshold swings (as low as 100 mV/dec), and minimal interfacial trap densities. Our phlogopite-based MoS$_2$ phototransistors achieve extraordinary optoelectronic performance, exhibiting responsivities reaching $3.3\times10^4$ AW$^{-1}$, photogain up to $6\times10^4$, and high detectivities (~$10^{10}$ Jones). These metrics highlight the intrinsic advantage of phlogopite in optoelectronic applications and its potential to advance high-performance multifunctional 2D electronics and photonics.

**RESULTS**

We begin by analysing the material properties of our phlogopite crystal. A representation of the atomic structure of a single phlogopite layer is shown in Fig. 1a (crystal structure adapted from ref. [25]). Phlogopite (KMg$_3$(AlSi$_3$)O$_{10}$(OH)$_2$) belongs to the mica family, a group of earth-abundant minerals and subset of phyllosilicates commonly found in the Earth's crust. Its structure is predominantly composed of oxygen, silicon, and aluminium, and it crystallizes in a monoclinic system. The material adopts a layered sheet-like morphology due to its characteristic T-O-T (tetrahedral-octahedral-tetrahedral) stacking structure[26–28]. In this 2:1 phyllosilicate framework, an octahedral sheet of cations is sandwiched between two sheets of linked (Si, Al)O$_4$ tetrahedra[29]. Compared to muscovite, another commonly studied mica, phlogopite is richer in magnesium and poorer in aluminium. Its colour can vary from golden brown (see Fig. 1a, inset) to reddish-brown depending on the presence of impurities such as iron (Fe), titanium (Ti), or fluorine (F), which can substitute for hydroxyl (OH⁻) groups. Increasing Fe content causes a transition toward biotite, while F enhances thermal stability[30]. The TOT structure promotes easy cleavage along the basal planes, where the mesh of basal oxygen atoms facilitates mechanical exfoliation of flakes with thicknesses ranging down to the ultrathin limit[31] (Fig. 1b).

We employ the conventional Scotch-tape method to exfoliate phlogopite onto PDMS (Gel-Pak) substrates, followed by a standard dry-transfer technique[32] to relocate the flakes onto desired substrates. The weak van der Waals forces between the layers facilitate mechanical exfoliation of flakes down to the monolayer limit[22,33,34]. Flake thicknesses can be easily distinguished by colour appearance, resulting from differences in optical contrast on $SiO_2$ substrates[35–37]. Using micro-reflectance spectroscopy and applying Fresnel law equations, we can determine layer thickness with high accuracy directly on the substrate, as previously demonstrated for other mica materials[38]. Out of these measurements we can also determine the refractive index for phlogopite, yielding in 1.45, consistent with earlier reports[38]. To verify the accuracy of the optical contrast thickness determination we measured the topography of the flake of Fig. 1b with atomic force microscopy (AFM, Supplementary Information Fig. S1). To assess the elemental composition, we performed X-ray photoelectron spectroscopy (XPS) on exfoliated phlogopite flakes transferred onto $SiO_2$ (290 nm)/Si substrates. The resulting spectrum (Fig. 1c) confirms Mg and K as the dominant substituents and shows negligible Fe(2p) and no detectable Ti peaks, indicating very low impurity content. Trace amounts of fluorine (F1s) are detected and are expected to enhance thermal stability. The XPS results confirm that the elemental composition of phlogopite includes major constituents such as silicon (44.2 %) and oxygen (33.4 %) also present in the substrate, minor constituents like magnesium (8.9 %), aluminium (8.8 %), and potassium (2.4 %), and trace amounts of iron (0.7 %) and fluorine (1.7 %). Raman spectroscopy of bulk phlogopite (see Supplementary Information) yields vibrational modes in agreement with literature[39,40], further validating the crystal quality. To determine the optical bandgap, we performed UV-vis absorption spectroscopy on thin exfoliated phlogopite flakes (Fig. 1d). Plotting the data as a Tauc plot with a direct optical transition[41] reveals a bandgap of 4.8 eV. This value is slightly higher than those previously reported for natural phlogopite (3.6 eV)[22], and lower than that of synthetic fluorphlogopite (~7.8 eV)[42]. The difference likely stems from our sample's low Fe impurity content, as discussed in earlier works[22,42]. Notably, many high-κ dielectrics suffer from relatively low bandgaps (<4 eV), which reduce the tunneling barrier and lead to higher leakage currents[14,43,44]. The large bandgap of natural phlogopite, therefore, provides a favourable combination of low leakage and high dielectric strength, making it an attractive candidate for use in 2D optoelectronic devices.

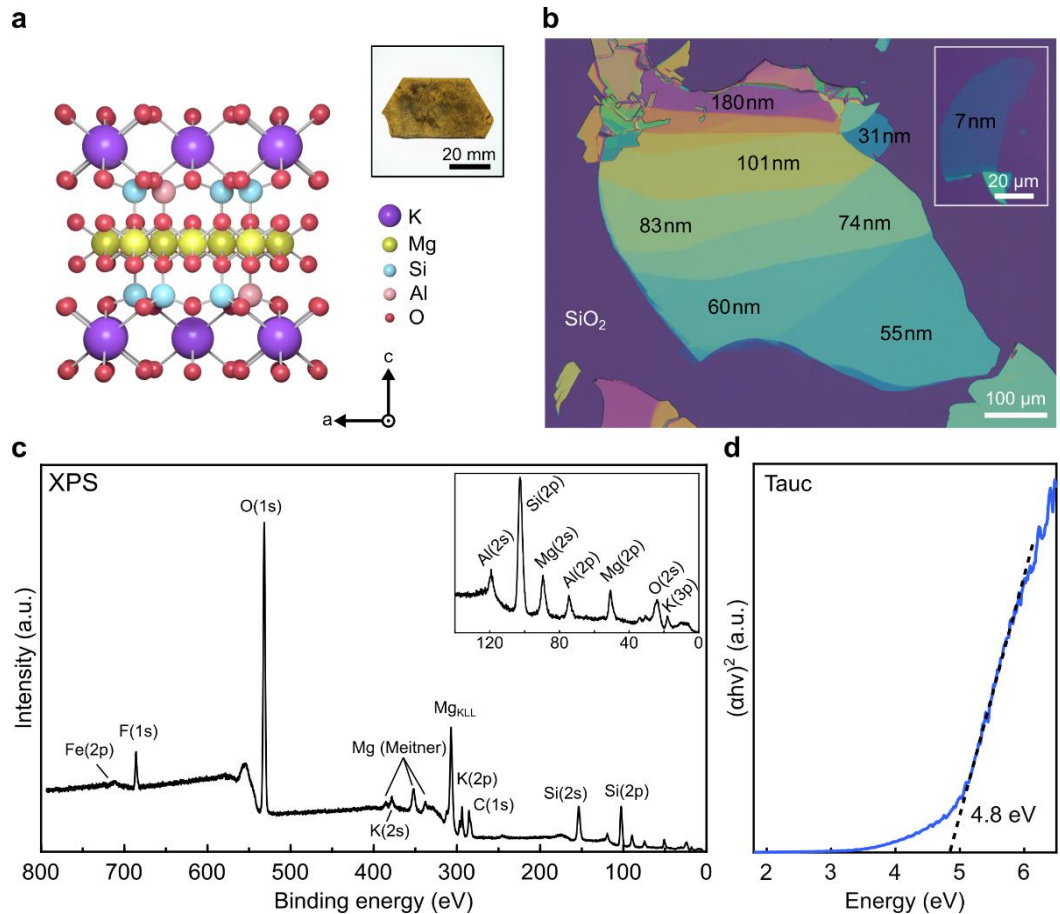

**Figure 1. Phlogopite material properties.** (a) Atomic structure of one layer phlogopite including crystal axes. (Inset) optical image of bulk crystal showing brown color. (b) Optical micrograph of layered phlogopite flakes, illustrating the rich optical contrast of different layer thicknesses down to 7 nm (inset) on 290 nm SiO$_2$/Si substrate. (c) X-ray photoemission spectroscopy (XPS) of phlogopite crystal. Mg and K represent the most prominent peaks. An almost diminishing iron peak Fe(2p) proofs that our crystal shows low impurity content. (d) UV-vis absorption spectroscopy of thin phlogopite layers represented by a Tauc plot of direct optical transition (α = 2), revealing an optical bandgap of 4.8 eV.

**Dielectric properties of thin phlogopite flakes**

Dielectric strength and insulating properties are critical in 2D optoelectronic applications, where selecting an appropriate dielectric layer can substantially influence device performance. Here, we use three complementary methods to characterize the dielectric constant of thin phlogopite flakes: conductive atomic force microscopy (C-AFM), double-gated field-effect transistor (FET) measurements, and traditional metal-insulator-metal (MIM) capacitance tests. Although MIM capacitance measurements can yield accurate results for sufficiently large and homogeneous flakes, fabricating large-area parallel-plate structures from ultrathin phlogopite is often challenging. Hence, we primarily rely on C-AFM and double-gated FET approaches for the thinner specimens.

First, C-AFM enables point-by-point measurement of tunneling current through phlogopite, allowing multiple measurements at different locations on the same flake. Additionally, the AFM gives information about surface roughness and contaminants. As shown in Fig. 2a–c, a 6.8 nm flake transferred onto a platinum electrode is scanned with a platinum AFM tip, where local current–voltage (I–V) sweeps are performed at multiple spots on the same flake. The resulting I–V curves can be fitted by a Schottky emission model, which directly relates the slope of the tunneling current to the dielectric constant[45,46] (Fig. 2c and more details on the model and fit in the Supplementary Information). From 13 measurement spots, we obtain an average dielectric constant of 10.7. Simultaneously, this approach provides insights into phlogopite's dielectric strength by monitoring I–V sweeps until breakdown, yielding a breakdown field of ~11.3 MV cm$^{-1}$. This value is comparable to those reported for other micas[21,23] and significantly surpasses those of newly proposed dielectric materials for 2D electronics[12,47,48]. Moreover, it exceeds the 1 V nm$^{-1}$ minimum breakdown field set by the International Roadmap for Devices and Systems (IRDS)[49]. Breakdown measurements on flakes of 6.8 nm and 22 nm thickness and more details on extracting the dielectric constant from the I-V slopes are presented in the Supplementary Information.

Next, we employ double-gated FET structures to validate these findings. A monolayer MoS$_2$ FET is fabricated on a SiO$_2$(290 nm)/Si substrate, then capped with a 35 nm phlogopite flake and lastly top-gated with a graphite flake (Fig. 2e). By recording the shift of the top-gate threshold voltage (V$_T$) under different back-gate biases, and modeling both gates as parallel-plate capacitors, we can extract the dielectric constant of phlogopite from the slope of the shift of $V_T$ (inset Fig. 2f), by using[15,50,51]

$$\frac{C_{SiO_2}}{C_P} = \frac{\varepsilon_{SiO_2} \cdot t_P}{\varepsilon_P \cdot t_{SiO_2}}, \qquad (1)$$

where $\varepsilon_{SiO_2}$ and $\varepsilon_P$ are the dielectric constants of $SiO_2$ (3.9) and phlogopite, respectively and $t_{SiO_2}$ and $t_P$ are their thicknesses. Specifically, the ratio of the top-gate to bottom-gate capacitances leads to a dielectric constant of 11.4 (Fig. 2f), which closely agrees with the C-AFM-based measurements.

Lastly, we conduct conventional MIM capacitance experiments for thicker flakes (50 nm and above), where uniform large area flakes are easier to achieve (Figs. 2g–i). We transfer 500×500 μm flakes onto prepatterned Au electrodes, then pattern and evaporate top Au contacts via maskless lithography to form parallel-plate capacitors. Measuring the capacitance at 100 Hz for four different flake thicknesses up to 160 nm results in an average dielectric constant of 12.5 (Fig. 2i). Overall, the dielectric constants of 10.7–12.5 place phlogopite well above that of hBN (~3)[52] and comparable to well-established materials such as $Si_3N_4$ or $Al_2O_3$[43]. Together with the electric breakdown field we compare phlogopite with other proposed dielectric materials for two-dimensional electronics in the Supplementary Information (Figure S4).

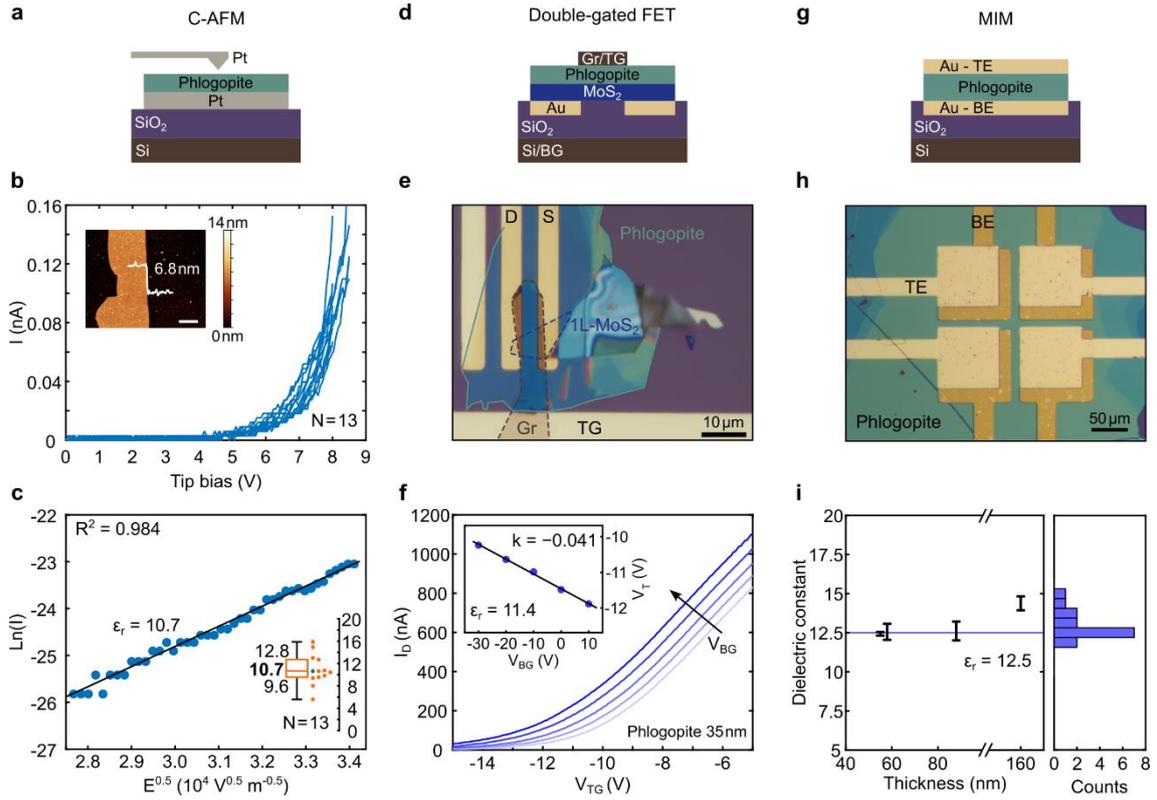

**Figure 2. Dielectric properties of layered phlogopite by three different methods.** (a)-(c) Determination of ultrathin layers of phlogopite by conductive atomic force microscopy. Schottky emission of a flake with a thickness of 6.8 nm was measured at multiple spots of the flake (b). The scale bar of the inset is 2 μm. All curves were fitted to the Schottky emission model and a mean dielectric constant of 10.7 was extracted (c). (d)-(f) Top- and bottom-gated Gr/Phlogopite/MoS$_2$/SiO$_2$ heterostructure for dielectric constant extraction. The transistor's threshold voltage ($V_T$) shift is recorded by sweeping the top-gate voltage for different back-gate biases (f). Extracting the slope k under a linear fit (f, inset) gives the dielectric constant by relating top (phlogopite) and bottom capacitances (SiO$_2$). (g)-(i) Metal-insulator-metal (MIM) capacitance measurements. Pre-patterned buried gold pads serve as bottom electrodes (BE), where large phlogopite flakes are transferred on top. Top gold electrodes (TE) are patterned using a maskless photolithography process to form a MIM structure. Capacitances are evaluated for four different thicknesses at f = 100 Hz, with four devices for each thickness. Dielectric constants for those thicknesses (extracted using the parallel plate capacitor model) are shown in (i).

**Natural phlogopite as insulator for 2D electronics**

To fabricate two-dimensional (2D) phototransistors, we integrate mechanically exfoliated monolayer MoS$_2$ with phlogopite as a gate insulator. We employ buried bottom-gated electrodes on SiO$_2$(290 nm)/Si(p$^{++}$) wafers, as illustrated by the three-dimensional schematic in the inset of Fig. 3a. The electrodes are patterned via maskless lithography and formed by a double-layer resist process to create undercut resist profiles. Applying a glass-etching cream to the exposed regions of $SiO_2$ produces 50 nm trenches, into which 5 nm Cr/45 nm Au is thermally evaporated, resulting in a planar surface between the electrodes and the $SiO_2$ substrate (see Ref. [53] for more fabrication details). This approach yields prepatterned substrates optimized for transferring 2D materials while minimizing mechanical stress.

As a next step, an 11 nm-thick flake of phlogopite is transferred onto the substrate to isolate the central gate electrode (G). A monolayer of $MoS_2$ is then transferred to span the gap between the source (S) and drain (D) contacts, as depicted in Fig. 3a. Figure 3b shows the gate-dependent current–voltage (I–V) characteristics for $V_{DS}$ up to 2 V, which exhibit current saturation. At small $V_{DS}$ (inset), the linear behaviour suggests low Schottky barriers[54]. Forward transfer characteristics of the field-effect transistor (FET) for different $V_{DS}$ values are displayed in Fig. 3c. These reveal a steep subthreshold swing (SS) of 140 mV dec$^{-1}$, an on/off current ratio of $10^5$, and reproducibility at various source–drain voltages. All measurements are performed at room temperature, with a transfer curve sweep rate of 0.02 Vs$^{-1}$. Notably, the device exhibits minimal hysteresis (~60 mV; see inset of Fig. 3c), especially when considering that phyllosilicate-based dielectrics commonly display higher hysteresis due to impurities[24,34] or water intercalation[55]. We attribute the comparatively small hysteresis to the low iron impurity content in our phlogopite crystal and the thermal annealing step preceding the measurements. Furthermore, using phlogopite flakes thinner than ~10 nm is critical for strong gate control. From the transfer curves, the estimated field-effect mobility is 6 cm$^2$V$^{-1}$s$^{-1}$ (see Supplementary Information for the transconductance curves and complete hysteresis data at various $V_{SD}$).

The density of interface traps ($D_{it}$) at the phlogopite/$MoS_2$ interface is evaluated via the standard expression[51]:

$$SS = \ln(10)\frac{k_B T}{q}\left(1 + \frac{qD_{it}}{C_{ox}}\right) \qquad (2)$$

where $k_B$ is the Boltzmann constant, T is the temperature, q is the elementary charge and $C_{ox}$ is the gate dielectric capacitance per unit area ($C_{ox}$ = 0.9 µF/cm$^2$ for phlogopite with a thickness of 11 nm). We obtain a $D_{it}$ of 7.6×10$^{12}$ cm$^{-2}$eV$^{-1}$, comparable to typical values reported for MoS$_2$ with conventional oxide dielectrics.[11]

To test the buried electrode design further, we fabricate a top-gated FET on monolayer MoS$_2$ with a reduced channel length of 3 µm. A gold top gate is transferred using a PDMS-based van der Waals pickup method. This top-gated device exhibits the same performance metrics (SS = 140 mV dec$^{-1}$) as the device on buried electrodes. Switching to a bilayer MoS$_2$ channel with 10 nm of phlogopite on buried electrodes improves the subthreshold slope to ~100 mV dec$^{-1}$, reduces $D_{it}$ to ~4.2×10$^{12}$ cm$^{-2}$eV$^{-1}$, and lowers hysteresis to ~30 mV (see Supplementary Information, Fig. S6). Such improvements are in line with reports showing that multilayer MoS$_2$ FETs often achieve lower SS[20,51,56].

Finally, we demonstrate the practicality of our buried electrode substrates and phlogopite/MoS$_2$ heterostructures by constructing an NMOS logic inverter with enhancement load (Fig. 3d–f). We prepare a substrate containing five buried electrodes, transfer the phlogopite and MoS$_2$ flakes, and arrange them so that both transistors share the same phlogopite and monolayer MoS$_2$. The MoS$_2$ is then laser-patterned into two separate channels (dashed black lines) without damaging the underlying phlogopite[57]. This "laser trimming" technique allows maximizing the geometry factor (the aspect ratio between the widths of the driver and load transistor), thereby increasing the slope of the inverter's voltage transfer characteristic. Figure 3e displays the inverter's output for different $V_{DD}$ values; the slight drop in output voltage compared to $V_{DD}$ stems from the enhancement load design, an effect that can be mitigated by using a depletion load inverter[19]. Nonetheless, the inverter maintains robust voltage gain, as shown in Fig. 3f. The flexibility of defining geometry factors is further illustrated in the Supplementary Information, where a different laser cut yields a distinct aspect ratio, demonstrating the adaptability of this approach for 2D logic circuits.

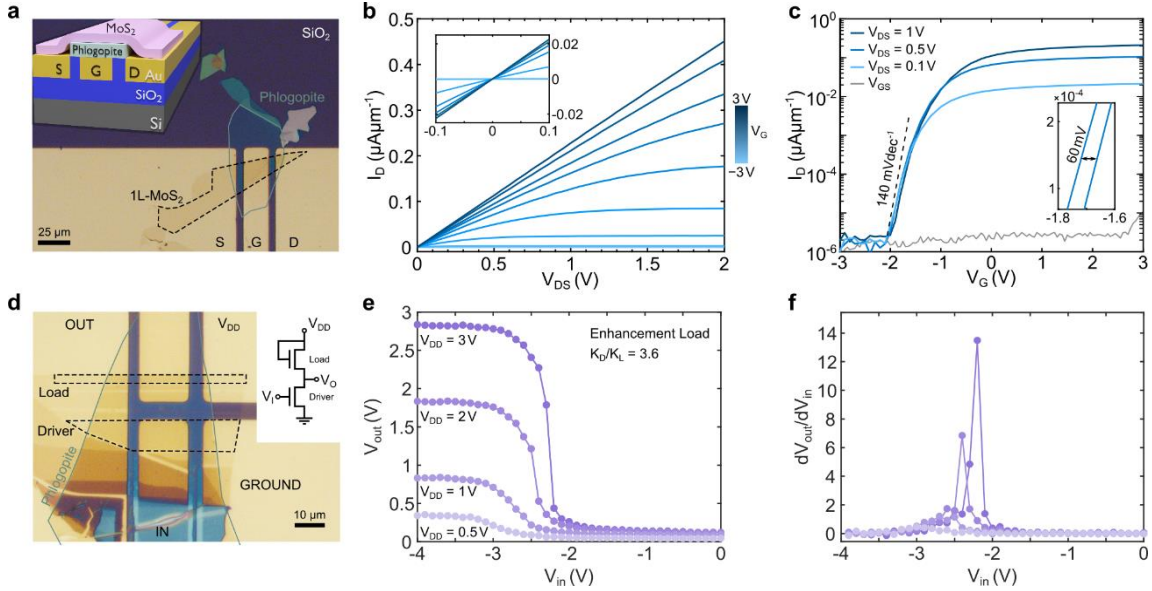

**Figure 3. Monolayer MoS$_2$ transistor and digital circuit inverter with phlogopite dielectric.** (a) Optical micrograph of single-layer MoS$_2$ transistor on buried gold electrodes with 11 nm thick phlogopite. The inset shows a 3D sketch of the device architecture. (b) Gate-dependent source-drain current-voltage (IVs) curves showing current saturation at higher V$_{DS}$ and linear behaviour at low V$_{DS}$ (inset). (c) FET forward transfer characteristics for different bias voltages and gate leakage. The inset is showing hysteresis of forward vs. backward sweep for V$_{DS}$ = 0.5 V. (d) Optical micrograph of single-layer MoS$_2$ NMOS inverter structure with enhancement load. The inset illustrates the analog circuit design. All electrodes are again buried and therefore planar with the SiO$_2$ surface. Both transistors use the same phlogopite and MoS$_2$ flake by design. The MoS$_2$ was laser-patterned with a 532 nm laser of a Raman system to define an enhancement load ratio of around 4 (ratio between the widths of the two transistors). (e) Inverter characteristics for different V$_{DD}$ voltages. A small voltage drop in comparison to V$_{DD}$ is natural due to the enhancement load of the circuit. (f) Inverter gain extracted from (e).

**Highly responsive 2D photodetectors**

In the following we evaluate the photoresponse characteristics of the MoS$_2$/phlogopite phototransistors. The optoelectronic performance of a monolayer MoS$_2$ device fabricated on a 29 nm-thick phlogopite flake is summarized in Figure 4. An optical micrograph of the device, AFM thickness determination of the phlogopite flake, and a differential reflectance spectrum of the monolayer $MoS_2$ before transfer are provided in the Supplementary Information.

Figure 4a shows the pre- and during-illumination transfer curves of the device (under a source-drain bias of $V_{SD} = 500$ mV and a wavelength of 660 nm), as well as the gate-dependent photocurrent defined as the difference between the two. Plotting the difference of the two curves allows to identify the gate bias with the highest photoresponse ($V_G = -8$ V). The device's spectral response in the visible is illustrated in Fig. 4b, extracting the photoresponsivity (R) from the measured photocurrent. R is defined as the amount of photocurrent generated per optical input power, by

$$R = \frac{I_p \cdot A_L}{P \cdot A_F} \qquad (3)$$

where $I_p$ is the detected photocurrent (inset of Fig. 4b), $A_L$ is the total illumination area (derived from the spot diameter), P is the illumination power and $A_F$ is the effective device area. The responsivity plot shows three peaks corresponding to the well-known A (660 nm), B (610 nm), and C (430 nm) excitonic transitions of $MoS_2$.

We focus on the device performance at the A exciton wavelength (660 nm) for detailed characterization. Fig. 4c shows the gate-dependent R for a fixed light power. As expected, the results follow very closely the shape of the transfer curve difference of Fig. 4a: the responsivity is maximized at a gate bias of –8 V and decreases the more the phototransistor is turned off. By setting the gate bias for maximal responsivity, we can further analyse the device's response for decreasing illumination power (Fig. 4d). Employing neutral density (ND) filters allows us to further increase R to a value of $3.3\times10^4$ AW$^{-1}$. This value is exceptionally high compared to previously reported MoS$_2$-based phototransistors, even considering devices operating at substantially higher source-drain currents[7,58,59].

From responsivity measurements, we calculate the photogain (G) assuming an external quantum efficiency of unity:

$$G = \frac{R \cdot h\upsilon}{q} \tag{4}$$

where h is Planck's constant, $\upsilon$ is the frequency of the incident light and q the elemental charge. The calculated maximum G is as high as $6\times10^4$, typically achievable only in hybrid structures or under significantly higher gate bias[58,60–62].

Increasing the negative gate voltage reduces the photogating effect, shifting the device's conduction mechanism to photoconductivity, thereby enabling faster detection speeds.[63] To illustrate this trade-off, we measure the rise time (t$_{rise}$, defined as the time interval for the signal to increase from 10% to 90% of its final value) at different gate biases and correlate it with responsivity (Fig. 4e). Responsivity reduction is accompanied by a substantially improved rise time down to 15 ms (fall time t$_{fall}$ = 120 ms) at a gate voltage of –20 V, demonstrating dynamic tuning capability.

Naturally as responsivity increases, also dark current increases and response times get longer. On the other hand, a way of identifying the smallest detectable signal of a photodetector is by determining its specific detectivity (D$^*$). We first need to look into the smallest optical power our device can detect, known as the noise equivalent power (NEP). There are two ways of determining the NEP. First the NEP can be derived from the power spectral density (PSD) of the photodetector's noise current and the responsivity, by $NEP = PSD/R$. However, as R can depend on the optical power, it is suggested to determine NEP experimentally[44]. We determine the NEP of our device at $V_G = -17$ V by acquiring power-dependent photocurrent with a modulated optical signal at 100 mHz. By performing a fast Fourier transform (FFT) of our measured current output, we can illustrate the power spectral density at each illumination power and extract the signal-to-noise ratio[43] (SNR, Figure 4f). The smallest detectable input power above an SNR of 1 is measured as 0.5 nW. The specific detectivity, considering bandwidth (f) and geometry (A, effective area of the device), is defined as:

$$D^* = \frac{\sqrt{A \cdot f}}{NEP} \tag{5}$$

Using the experimentally obtained NEP we calculate a specific detectivity of $D^*_{100mHz} = 1.5 \times 10^8$ J. To compare, we can calculate D* from the dark current PSD ($8 \times 10^{-11}$ AHz$^{-1/2}$, see Supplementary Information) and the responsivity (200 AW$^{-1}$, at $V_G = -17$ V, see Fig. 4c), obtaining $2.16 \times 10^8$ J at 100 mHz. The values of the two methods match well, as the responsivity has been acquired at low power density, close to SNR = 1. Therefore, we can determine the specific detectivity at 1 Hz, which is the preferred value for benchmarking purposes, using the second method. With a PSD of $5 \times 10^{-12}$ AHz$^{-1/2}$ the detectivity was found to be $D^*_{1Hz} = 1.5 \times 10^{10}$ J. At low power densities the photocurrent behaves according to a power law of $I_P \approx P^\alpha$, with α = 0.88. See Table S1 in the Supplementary Information for a benchmarking table comparing our phlogopite/MoS$_2$ phototransistor with other MoS$_2$ photodetectors of various dielectric materials.

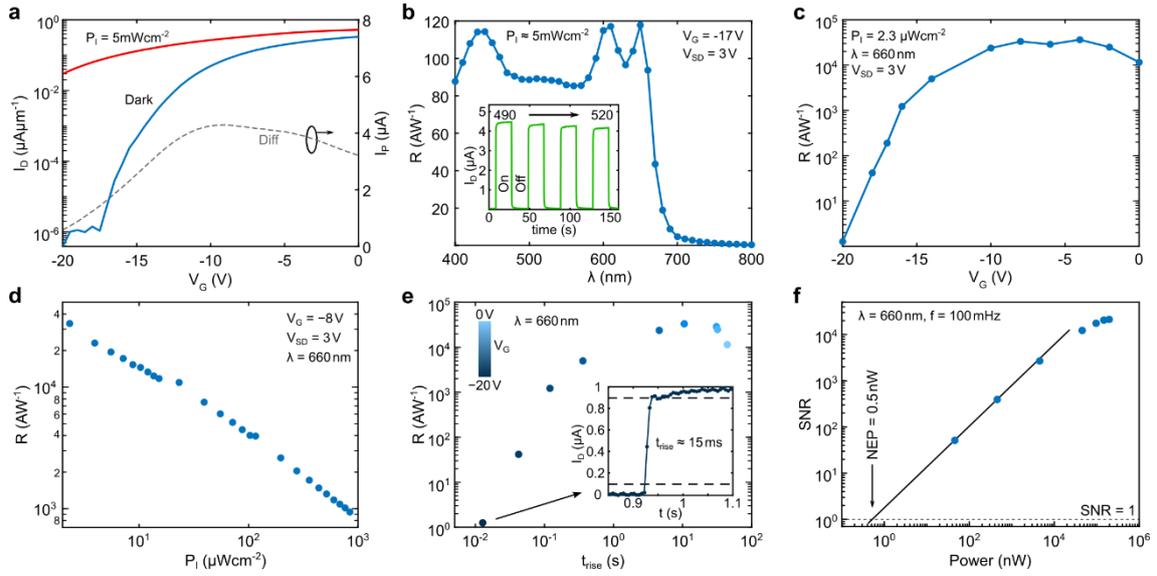

**Figure 4. Monolayer MoS$_2$ photodetector with phlogopite dielectric.** (a) Pre- and during-illumination transfer characteristics of a 1L-MoS$_2$ phototransistor under V$_{SD}$ = 500 mV after exposure with a 660 nm LED with P$_i$ = 5 mWcm$^{-2}$. The panel also plots the difference between the two curves (dashed grey line), identifying the gate voltage with highest photo response (V$_G$ = –8 V). The phlogopite has a thickness of 30 nm. (b) Wavelength-dependent photoresponsivity under V$_G$ = –17 V and V$_{SD}$ = 3 V. The inset shows selected bias current vs. time pulses, from which the responsivity curve is extracted. (c) Gate voltage-dependent responsivity at 660 nm for a fixed illumination power of 2.3 μWcm$^{-2}$. (d) Illumination power-dependent photoresponsivity for a fixed gate (V$_G$ = –8 V) and source-drain bias (V$_{SD}$ = 3 V), with a maximum responsivity of 3×10$^4$ AW$^{-1}$. (e) Response time vs. responsivity for different gate bias for a fixed illumination wavelength of 660 nm. Gate bias is depicted by colour intensity. The inset shows the pulse in the I-t measurement corresponding to the fastest response time of the device (15 ms). (f) Experimental NEP determination for a fixed wavelength and gate bias (V$_G$ = –17 V). Data points correspond to signal-to-noise ratios (SNR) extracted by a Fourier transform of the measured photocurrent for decreasing light power from pulsed illumination measurements at f = 100 mHz. The intersection between the linear fit with a SNR of 1, gives the NEP for the device, from which detectivity can be calculated.

To confirm these results, an additional monolayer MoS$_2$ phototransistor is fabricated and characterized, showing comparable behavior with a maximal photoresponsivity of 1×10$^4$ AW$^{-1}$ (see Supplementary information).

Lastly, we evaluate phlogopite as a dielectric with another 2D semiconductor, WS$_2$ (see Figure S11, Supplementary Information). A phototransistor comprising single-layer WS$_2$ on a 32 nm phlogopite flake similarly exhibits distinct A, B, and C exciton peaks, reaching a maximal responsivity of 22 AW$^{-1}$ at 617 nm illumination, surpassing commonly reported values for WS$_2$-based devices[64–66].

**CONCLUSIONS**

In summary, we have demonstrated that naturally occurring phlogopite mica, when exfoliated to the ultrathin limit, functions as a high-quality gate dielectric for monolayer and few-layer TMD transistors and photodetectors. The phlogopite flakes exhibit a wide bandgap (~4.8 eV), a dielectric constant of ~11, and a high breakdown strength (>10 MVcm$^{-1}$), satisfying key criteria for low-power (opto-)electronics. By directly integrating phlogopite flakes with MoS$_2$, we realized phototransistors with low subthreshold swings (down to 100 mVdec$^{-1}$), achieve responsivities up to $3.3\times10^4$ AW$^{-1}$ and detectivities approaching $10^{10}$ Jones. Moreover, we can effectively tune the photo response characteristics of our devices by gate bias. These findings position phlogopite as a sustainable and readily available gate dielectric for advanced 2D electronic and optoelectronic applications, providing new opportunities to push device performance, reduce power consumption and enable damage-free integration with various 2D channels.

MATERIALS AND METHODS

**Materials, exfoliation and transfer**

$MoS_2$ flakes were mechanically exfoliated from a natural molybdenite mineral (Molly Hill Mine, Quebec, Canada) using Nitto tape (Nitto SPV 224) and phlogopite flakes from its bulk crystal (Madagascar) using Scotch tape (Magic tape by 3M) onto a PDMS substrate (Gel-Film WF 4 × 6.0 mil by Gel-Pack). Phlogopite flakes were transferred to a $SiO_2$ (290 nm)/Si (p++) substrate, identified and stamped to the final substrate using common nail polish transfer[67]. $MoS_2$ flakes were identified on PDMS under an optical microscope[68] and dry-transferred to the final substrate[32].

**XPS characterization**

XPS measurements of phlogopite flakes on $SiO_2$ (290 nm)/Si substrate were carried out in ANA chamber (STARDUST machine UHV module)[69] using a Phoibos 100/150 electron/ion analyser with a 1D-DL43 2-100 1-dimensional delay line detector and XR50 M X-ray source Aluminium anode (Al K$\alpha$ =1486.7 eV), the X-rays were passed through a FOCUS 500/600 monochromator. The analyser entrance and exit slits were 7 x 20 mm and 'open', respectively.

**Buried electrode fabrication**

Buried electrodes were patterned on a $SiO_2$ (290 nm)/Si (p++) wafer by mask-less lithography (Microlight3D SmartPrint), a 50 nm etch of the $SiO_2$ layer (glass etching cream Armour Etch) using the photoresist as a mask, and subsequent thermal evaporation of 5 nm Cr and 45 nm Au into the etched trenches[53,70].

**C-AFM characterization**

A commercial Atomic Force Microscopy (AFM) system, from Nanotec, operating in ambient conditions was employed to perform morphological and conductive characterization of the sample. Measurements have been acquired in dynamic (topography) and in contact modes (C-AFM) using commercial tips from Nanosensors (PPP-FMR) and Rocky Mountain Nanotechnology (25Pt300B) respectively. Current - Voltage (I-V) curves were acquired at different locations in the sample, while maintaining a contact force between 200-500 nN. During the measurements, a DC bias (ranging from 0 to 9 V) was applied to the tip while the sample was grounded.

**Electrical characterization**

Electrical characterizations were carried out in a home-built measurement setup at room temperature under vacuum conditions ($10^{-6}$ mbar), where the samples can be in-situ annealed at 200°C for 2h.[71] For FET measurements a source-meter unit (Keithley 2450) was used to perform source-drain sweeps and two programmable benchtop power supplies (Tenma, model 72–2715) are connected in series to perform gate voltage sweeps. For four-probe measurements (double-gated FET (Fig.2d-f), NMOS inverter (Fig. 3d-f)) an additional benchtop power supply was used.

**Optoelectronic characterization**

The photoresponse of our devices was tested in the same vacuum chamber as the electrical measurements. We use multimode fiber-coupled light sources for optical illumination inside the vacuum chamber. The fiber-coupled light is guided through a tube lens system onto our device, resulting in a circular spot of 900 μm in diameter with homogeneous power density over the sample.[72,73] A tunable xenon lamp source (Bentham TLS120Xe) was used to investigate wavelength-dependent photoresponse. A 660 nm LED source (Thorlabs, MxxxFy series) in combination with ND filters (Thorlabs NEK01S) enabled us to study power-dependent photocurrent generation over a wide range of illumination power.

## DATA AVAILABILITY

The datasets generated during and/or analysed during the current study are available from the corresponding author on reasonable request.

## CODE AVAILABILITY

The codes used for plotting the data are available from the corresponding authors on reasonable request.

## ACKNOWLEDGEMENTS

The authors would like to thank Beatriz H. Juárez for her help on the UV-vis absorption measurements. We are also thankful to the Geomineral Museum of Madrid (Museo Geominero) for the organization of the monthly mineral market. We acknowledge funding from the Spanish Ministry of Science and Innovation (Grants PID2020-115566RB-I00, TED2021-132267B-I00, PDC2023-145920-I00, PRE2022-105538 and PID2023-149077OB-C31) and the Comunidad de Madrid (Project TEC-2024/TEC-308).

## AUTHOR INFORMATION

**Authors and Affiliations**

**2D Foundry Research Group. Instituto de Ciencia de Materiales de Madrid (ICMM-CSIC), Madrid, Spain**

Thomas Pucher, Julia Hernandez Ruiz, Carmen Munuera and Andres Castellanos-Gomez

**ESISNA Research Group. Instituto de Ciencia de Materiales de Madrid (ICMM-CSIC), Madrid, Spain**

Guillermo Tajuelo-Castilla, José Ángel Martín-Gago

**Author Contributions**

T.P. and A.C.-G. conceived the idea and designed the experiments. T.P. fabricated the devices and performed the (opto-)electronic measurements and analysis. J.H.R. and C.M. performed the AFM and C-AFM characterizations. G.T.-C. and J.A.M-G. performed the XPS measurements. T.P. wrote the first draft of the paper. The manuscript was written through contributions of all authors. All authors have given approval to the final version of the manuscript.

**Corresponding Authors**

Thomas Pucher thomas.pucher@csic.es

Andres Castellanos-Gomez andres.castellanos@csic.es

**ETHICS DECLARATIONS**

**Conflict of interest**

The authors declare no conflict of interest.

# Supplementary Information:

# Natural layered phlogopite dielectric for ultrathin two-dimensional optoelectronics

*Thomas Pucher[1*], Julia Hernandez-Ruiz[1], Guillermo Tajuelo-Castilla[2], José Ángel Martín-Gago[2], Carmen Munuera[1] and Andres Castellanos-Gomez[1*]*

*[1]2D Foundry Research Group. Instituto de Ciencia de Materiales de Madrid (ICMM-CSIC), Madrid, E-28049, Spain.*

*[2]ESISNA Research Group. Instituto de Ciencia de Materiales de Madrid (ICMM-CSIC), Madrid, E-28049, Spain.*

thomas.pucher@csic.es

andres.castellanos@csic.es

**Optical thickness determination of phlogopite flakes**

The apparent color of phlogopite flakes, which depends on their thickness, can be understood using a multi-layer optical model, shown in previous works [S1,S2]. In this model, incident light strikes a surface—either the substrate or the flake—and then passes through multiple layers until it reaches the silicon substrate (Fig. S1c). At this point, the light retraces its path (assuming normal incidence) and is ultimately detected by the spectrometer. The reflection and transmission of the light within this system can be determined using Fresnel's equations. The reflected intensity for monochromatic light under normal incidence is given by:

$$I(\lambda) = \left| \frac{r_{01}e^{i(\phi_1+\phi_2)} + r_{12}e^{-i(\phi_1-\phi_2)} + r_{23}e^{-i(\phi_1+\phi_2)} + r_{01}r_{12}r_{23}e^{i(\phi_1-\phi_2)}}{e^{i(\phi_1+\phi_2)} + r_{01}r_{12}e^{-i(\phi_1-\phi_2)} + r_{01}r_{23}e^{-i(\phi_1+\phi_2)} + r_{12}r_{23}e^{i(\phi_1-\phi_2)}} \right|^2$$

Where the reflection coefficient is $r_{ij} = (n_i - n_j)/(n_i + n_j)$, the phase shift in the medium is $\phi_i = 2\pi n_i d_i/\lambda$, with its refractive index $n_i$, its thickness $d_i$ and the wavelength $\lambda$. The indices 0,1,2 and 3 are standing for air, phlogopite, $SiO_2$, and Si, respectively.

The optical contrast can be obtained from

$$C = \frac{I_1 - I_0}{I_1 + I_0}$$

Using this model, in combination with our micro-reflectance setup [S3], we can obtain the flake thicknesses of Figure S1a, optically. The differential reflectance and optical contrast values are plotted in Figure S1c&d. By using AFM to resolve the topography of the flake, we can confirm the accuracy of the optical thickness determination method and determine the index of refraction of phlogopite with 1.45, in agreement with previous work [S4].

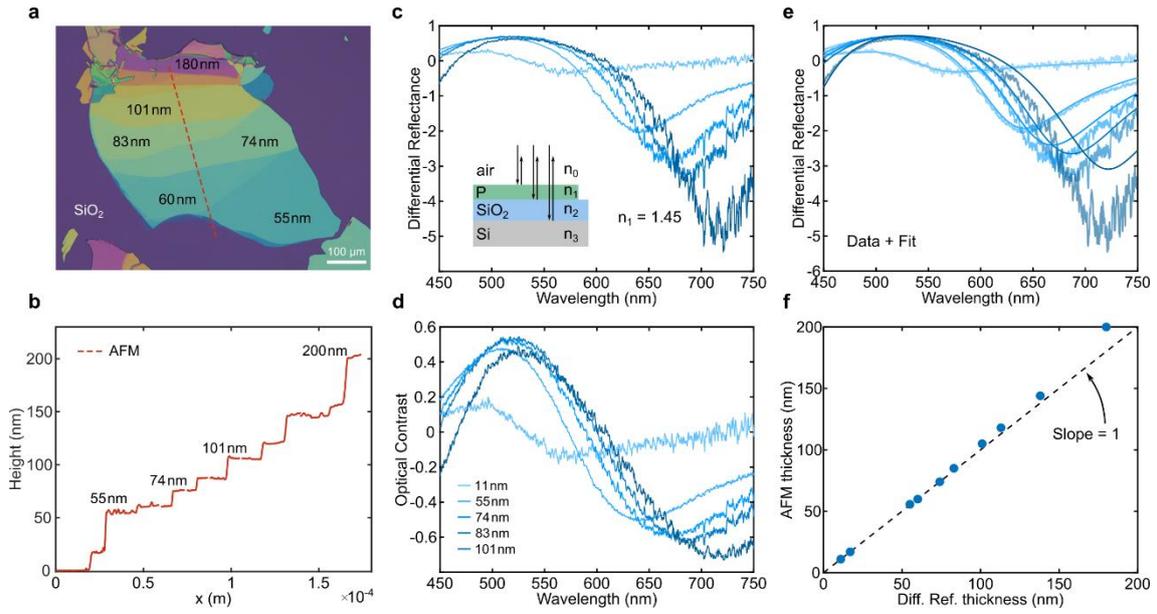

**Figure S1. Phlogopite thickness determination.** (a) Optical microscopy image of flake with different thicknesses presented in the main text. Indicated thicknesses have been measured optically through micro-reflectance measurements. The red dashed line marks the path of the AFM topography measurements of (b). Phlogopite layer terraces have been measured in individual maps and glued together for demonstration purposes. (c) Differential reflectance curves for selected thicknesses as function of the wavelength. The inset shows an illustration of the different layers involved in the Fresnel law equations, with corresponding refractive index numbers. (d) Optical contrast of the same thicknesses as in (c). The $SiO_2$ has a thickness of 290 nm. (e) Differential reflectance curves including the corresponding fits. (f) Comparison between thickness values obtained optically and the AFM. The dashed slope of 1 indicates a perfect fit. The optical thickness determination is vary accurate up to around 150 nm of thickness.

## Raman analysis of phlogopite

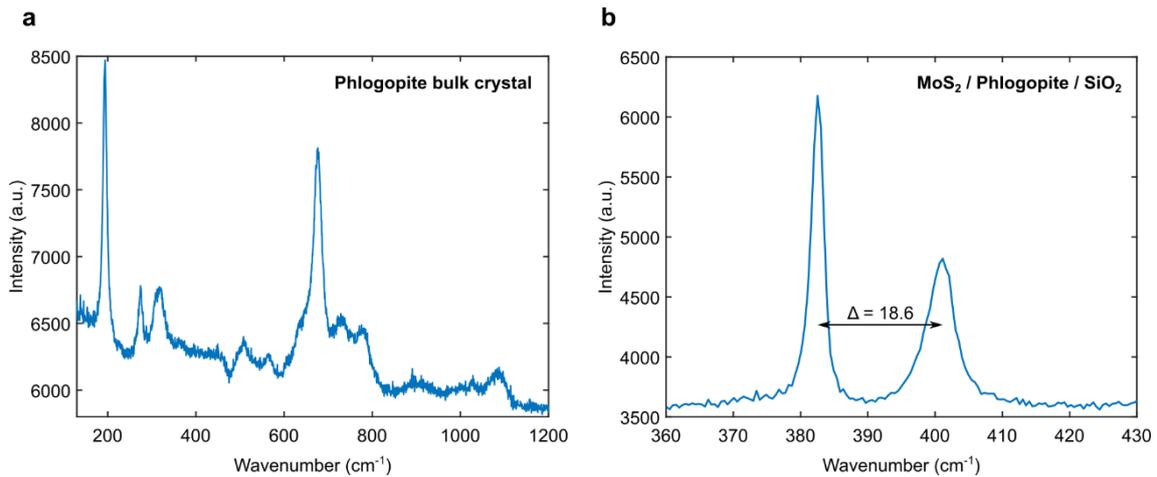

**Figure S2. Raman spectroscopy of phlogopite.** (a) Raman spectroscopy of bulk phlogopite crystal. (b) Raman spectroscopy of monolayer $MoS_2$ on top of a phlogopite flake on a $SiO_2/Si$ substrate.

**Dielectric constant determination by Schottky emission extraction**

Determining the dielectric constant of atomically thin nanosheets can be challenging due to their small lateral dimensions. Fabricating large-scale capacitors with such small samples becomes impractical due to difficulties in ensuring uniformity, reliable contact, and consistent layer thickness over the entire area. By using a conductive AFM tip to apply a voltage and measure the current through the nanosheet, it is possible to characterize the material's electrical properties at a local scale. When an electric field is applied across the nanosheet, the current can be analyzed using the Schottky emission mechanism to extract the dielectric constant of the material. [S26,S27]

Schottky emission or field-assisted thermionic emission, is described by the Richardson-Dushman equation, which states that the current (I) depends on the temperature (T) and the work function (φ):

$$I = AT^2 e^{\left(-\frac{\phi - \Delta\phi}{k_B T}\right)}$$

where $\Delta\phi = \sqrt{\frac{eE}{4\pi\varepsilon_0\varepsilon_r}}$ is the field-dependent reduction in barrier height (A is the Richardson constant, E is the electric field, $\varepsilon_0$ is the vacuum permittivity, $k_B$ is the Boltzmann's constant, and $\varepsilon_r$ is the relative permittivity). The electric field across the flake is related to the applied voltage (V) and the flake thickness (d), resulting in a modified Schottky emission equation:

$$I = AT^2 e^{\left(-\frac{\phi - \sqrt{\frac{eV}{4\pi\varepsilon_0\varepsilon_r d}}}{k_B T}\right)}$$

To extract the contribution of the relative permittivity, we linearize this equation taking the natural logarithm to

$$\ln(I) = \ln(AT^2) - \frac{\phi}{k_B T} + \frac{\sqrt{\frac{eV}{4\pi\varepsilon_0\varepsilon_r d}}}{k_B T}$$

Let $C = \sqrt{\frac{e}{4\pi\varepsilon_0 d}}$, the equation becomes

$$\ln(I) = \ln(AT^2) - \frac{\phi}{k_B T} + \frac{C\sqrt{V}}{\sqrt{\varepsilon_r} T}$$

When plotting $\ln(I)$ vs $\sqrt{V}$, the dielectric constant $\varepsilon_r$ can be directly extracted from the slope $k$ of the linear region.

$$\varepsilon_r = \left(\frac{C}{k \cdot T}\right)^2$$

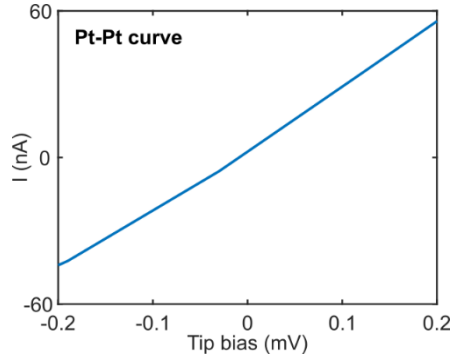

**Figure S3. Pt-Pt comparison curve of C-AFM.** The platinum tip is positioned on top of the platinum electrode without a flake in between to validate the electrical connection.

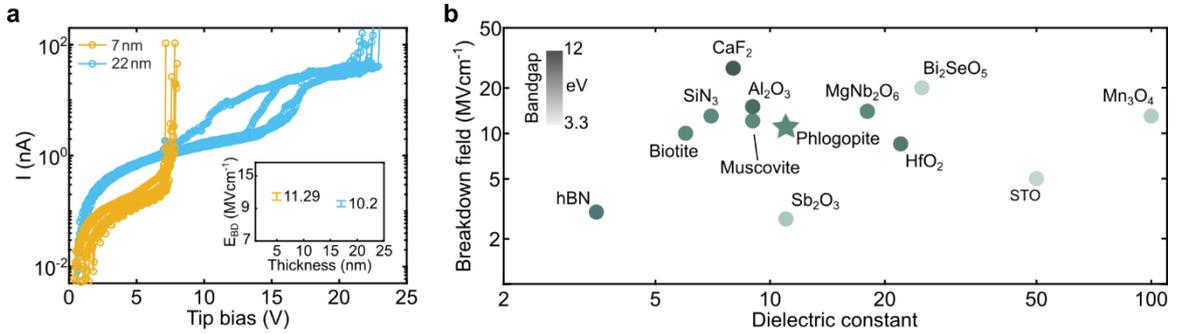

**Figure S4. Dielectric breakdown of phlogopite flakes and comparison with other dielectrics.** (a) Two flake thicknesses were further investigated under the C-AFM for dielectric breakdown (7 nm and 22 nm). Breakdown fields are indicated in the inset and exceed 1 V/nm. (b) Comparison of different dielectrics suggested for two-dimensional electronics. Bandgap size is indicated by color intensity. Information taken from Refs S5-S25.

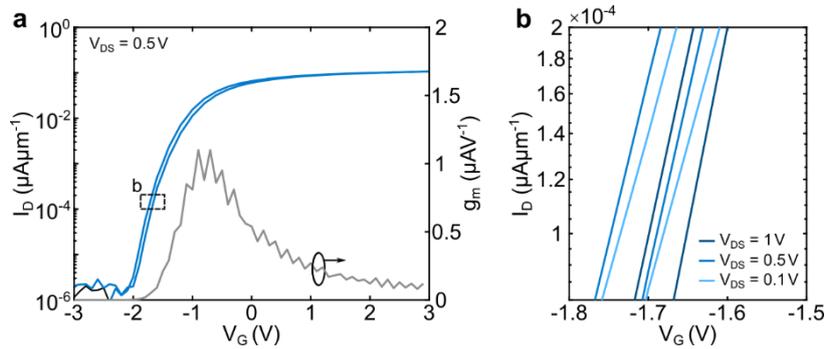

**Figure S5. Monolayer MoS2 FET details for the device in the main text.** (a) Full transfer curve and extracted transconductance for a $V_{DS}$ of 0.5 V. The maximum transconductance value is taken for field-effect mobility calculations. (b) Magnification of the subthreshold regime of all three transfer curves from the main text, illustrating the consistent hysteresis of ~60 mV even for varying $V_{DS}$. Sweep rate is 0.02 V/s and all measurements were performed under vacuum conditions (~e-6mbar) at room temperature.

## Bilayer MoS₂ FET

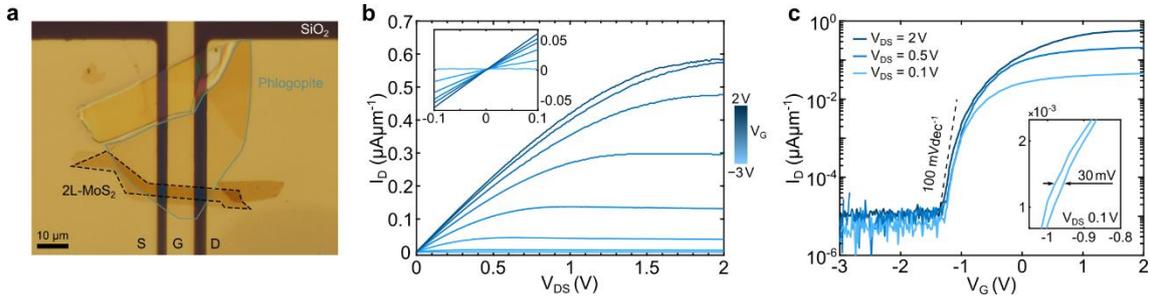

**Figure S6. Bilayer MoS₂ FET details.** (a) Optical micrograph of the bilayer FET with a phlogopite of 10 nm thickness. (b) Gate-dependent source-drain current-voltage (IVs) curves showing current saturation at higher $V_{DS}$ and linear behavior at low $V_{DS}$ (inset). (c) FET forward transfer characteristics for different bias voltages. The inset shows the hysteresis of the transfer curve for a source-drain bias of 0.1 V.

## Top-gated monolayer MoS₂ FET

We use pre-patterned (buried) electrodes to first transfer single-layer MoS₂ and then thin phlogopite on top. To realise the top-gate gold transfer, we first prepare gold strips of desired size on SiO₂/Si substrate by maskless lithography and thermal Au evaporation (50 nm), without evaporation of a sticking layer. Then we prepare a stamp consisting of a rectangular piece of PDMS (Gel-Film WF 4 × 6.0 mil by Gel-Pack) mounted on a glass slide, overhanging like a cantilever. We bring the PDMS in contact with the gold strip and peel it off as fast as possible to pick up the gold strip. As a final step the gold strip is transferred on top of the MoS₂/phlogopite stack by dry-transfer, acting as a top-gate electrode.

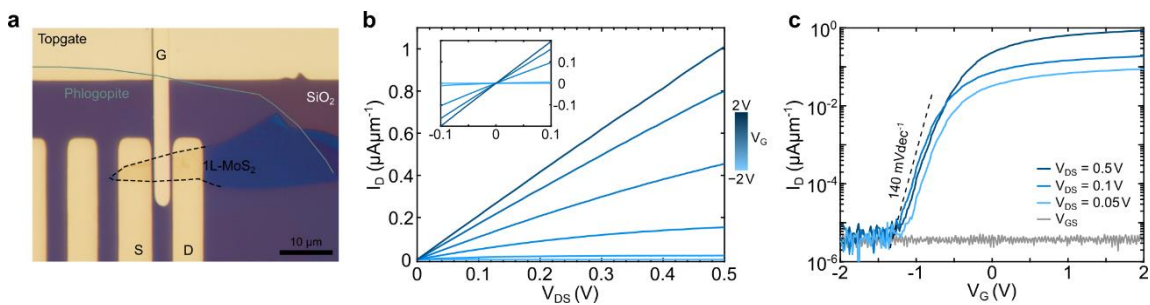

**Figure S7. Top-gated monolayer MoS₂ FET details.** (a) Optical micrograph of the top-gated FET. The phlogopite has a thickness of 9 nm. (b) Gate-dependent source-drain current-voltage (IVs) curves showing current saturation at higher $V_{DS}$ and linear behaviour at low $V_{DS}$ (inset). (c) FET forward transfer characteristics for different bias voltages and gate leakage.

**Monolayer MoS$_2$ inverter with different enhancement load**

NMOS inverters usually are categorized in inverters with resistor load, enhancement load or depletion load. Their input-output relationship is subject to the geometry factor between the load and driver transistor. The sharpness of the transition, as well as the minimum output voltage decrease with increasing load resistance. An inverter with enhancement load uses a load transistor instead of a resistor, decreasing its footprint dramatically. The two transistors in the inverter circuit are defined by their threshold voltage and their geometrical aspect ratio $K_D/K_L$, where $K_X$ is the transistor's width-to-length ratio. As $K_D/K_L$ increases, the input-output relationship of the inverter improves. [S28] Due to the design of our circuit, the lengths of both driver and load transistor is always the same, which means that the aspect ratio is solely defined by the width ratio of the two transistors, which we can define by laser cutting. Illustrating the importance of the aspect ratio, we show the inverter device of the main text after an initial laser cut that only separated the two transistors with one cut (panel a), resulting in an aspect ratio of 1.1. Input-output characteristics and inverter gain are shown below and prove poor device performance. After these measurements load and driver transistor have been laser cut again, to define an aspect ratio of around 4, which corresponds to the results shown in the main text.

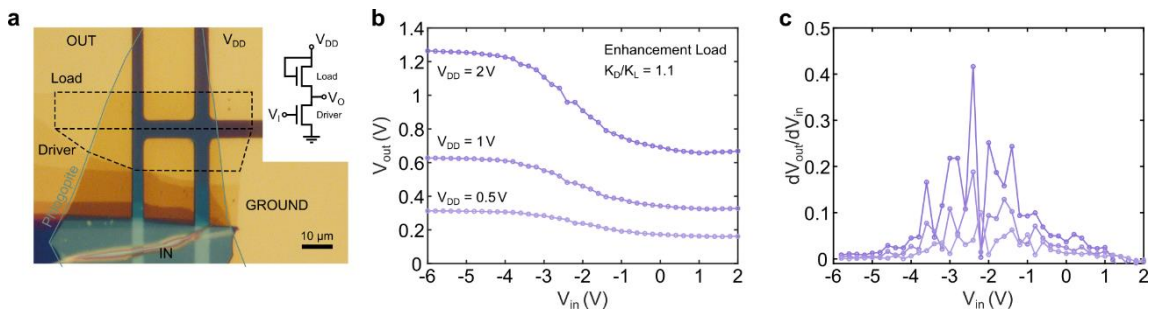

**Figure S8. Monolayer MoS2 inverter with different enhancement load.** (a) Optical micrograph of single-layer MoS$_2$ NMOS inverter structure with enhancement load factor of 1.1, before final laser cutting. (b) Inverter characteristics for different $V_{DD}$ voltages. Parasitic effects for this configuration are large. (c) Inverter gain extracted from (b).

**Optical microscope image, AFM and differential reflectance of MoS₂ phototransitor**

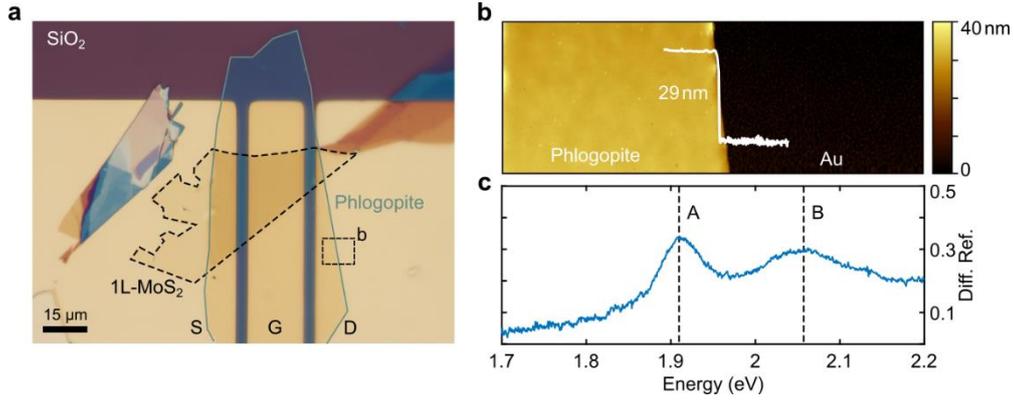

**Figure S9. Details of device of main figure 4.** (a) Optical micrograph of single-layer MoS$_2$ phototransistor on a 29 nm thick phlogopite flake. (b) AFM topography of the area marked in (a). (c) Differential reflectance spectroscopy of the MoS$_2$ flake on PDMS before transfer onto the phlogopite.

**Noise equivalent power (NEP) extraction and specific detectivity (D*) calculation**

The NEP is defined as the smallest detectable optical power of the photodetector at a certain bandwidth, whereas the detectivity D is the reciprocal of NEP:

$$D = \frac{1}{NEP}$$

The specific detectivity takes bandwidth (f) and geometry factors (A, effective area of the device) into account and is defined by:

$$D^* = \frac{\sqrt{A \cdot f}}{NEP}$$

To determine NEP without overestimating detectivity it is crucial to experimentally measure NEP and therefore the photodetector's noise. Calculating the noise simply by $PSD = \sqrt{2qI_{dark}f}$ can lead to a largely underestimated NEP value and therefore overestimated D*.

For example, if we calculate the detectivity of our system with this formula we get a value of $D^*_{1Hz} = 2 \times 10^{11}$ J, which is one order of magnitude higher than the value from the experimental NEP determination in the main text.

The current spectral density of the device of main Figure 4 is plotted in Figure S9a, including the instrumentation noise floor with lifted probe tips (labelled "open"). We see that 1/f noise (shot noise) is the dominating noise contribution at low frequencies.

To determine the NEP experimentally, we measure photocurrent while modulating the input LED light with a function generator at a frequency of 100 mHz. Doing so for decreasing power allows us to extract signal-to-noise (SNR, difference between the noise floor and the peak amplitude) at the modulation frequency by Fourier transforming the photocurrent measurement.

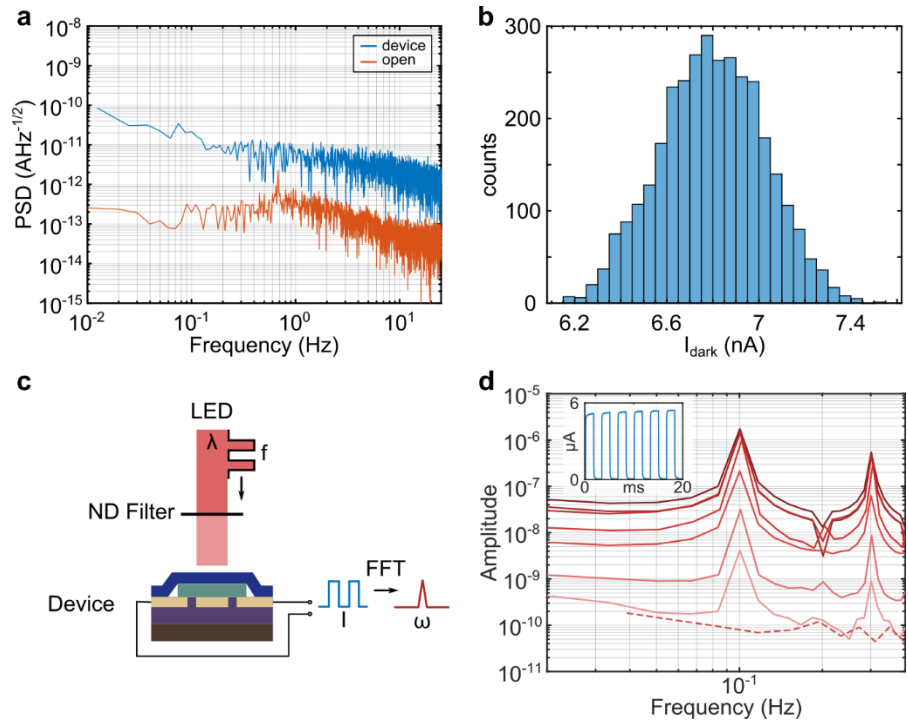

**Figure S9. MoS$_2$ phototransistor noise and NEP determination.** (a) Current spectral density of the dark current under a gate bias of $V_G$ = –17 V and $V_{SD}$ = 3 V (device) and with the probe tips lifted (open). (b) Histogram of dark current under the same conditions. (c) Schematic of experimental NEP determination. The light of the LED gets modulated by a function generator with the frequency f (in our case 100 mHz) and passes through neutral density (ND) filters onto the device. The modulated output current (I) is measured with a Keithley source-meter and Fourier transformed to extract the SNR. (d) Fourier transformed current signals (ω) around the modulation frequency. The noise floor (SNR = 1) is indicated by a dashed line. The inset shows the measured output current (I) for a configuration without ND filter. The first harmonic of the square signal appears at 0.3 Hz (3f).

## MoS$_2$ phototransistor benchmarks for various dielectric systems

The following table summarizes figures of merit for phototransistors in literature with the focus on photodetection in the visible red spectrum, comparable to our results.

| Material | λ (nm) | R (AW$^{-1}$) | D* (J) | t rise/fall (ms) | REF |
|---|---|---|---|---|---|
| SiO$_2$ | 400-600 | 880 | - | 4000/9000 | S29 |
| SiO$_2$ | 532 | 2200 | - | -/55 | S30 |
| Al$_2$O$_3$ | 625 | 1x10$^4$ | - | 10/0.1 | S31 |
| SNO | 600 | 163 | 3.8×10$^{14}$ * | - | S32 |
| HfO$_2$ | ~620 | 1.3×10$^4$ | - | - | S33 |
| SiO$_2$/MoS$_2$/HfO$_2$ | 635 | 5×10$^4$ | 7.7×10$^{11}$ | 7 | S34 |
| Si$_3$N$_4$ | 647 | ~200 | - | 13000/11000 | S35 |
| P(VDF-TrFE) | 635 | 2570 | 2.2×10$^{12}$ * | 10/10 | S36 |
| SiO$_2$ | 532 | 0.57 | 10$^{10}$ * | 0.07/0.11 | S37 |
| MoO$_x$ | 638 | 10$^{-2}$ | 10$^{10}$ * | 52/52 | S38 |
| Sb$_2$O$_3$ | 671 | 2×10$^4$ | 10$^{15}$ * | 60/55 | S39 |
| **Phlogopite** | **660** | **3.3×10$^4$** | **10$^{10}$** | **15/120** | **This work** |

**Table S1. MoS$_2$ phototransistor benchmark comparison.** Values for responsivity and detectivity are given as maximum device values, which mostly is not measured at the same gate bias (e. g. photo conduction mechanism).

Note: Specific detectivity values marked with * are not directly measured but calculated using the formula $D^* = R \frac{\sqrt{A}}{\sqrt{2e \cdot I_{dark}}}$ and therefore very likely overestimated, as previously explained.

## Additional MoS₂ photodetector device

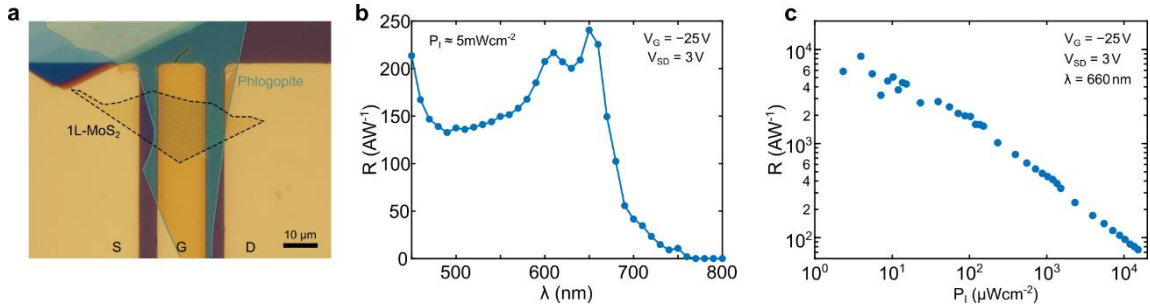

**Figure S10. Additional monolayer MoS₂ phototransistor details.** a) Optical micrograph of the device. The phlogopite has a thickness of 51 nm. (b) Wavelength-dependent photoresponsivity under $V_G$ = –25 V and $V_{SD}$ = 3 V. (c) Illumination power-dependent photoresponsivity for a fixed gate ($V_G$ = –25 V) and source-drain bias ($V_{SD}$ = 3 V), with a maximum responsivity of $1\times10^4$ AW$^{-1}$.

## Monolayer WS₂ phototransistor on phlogopite

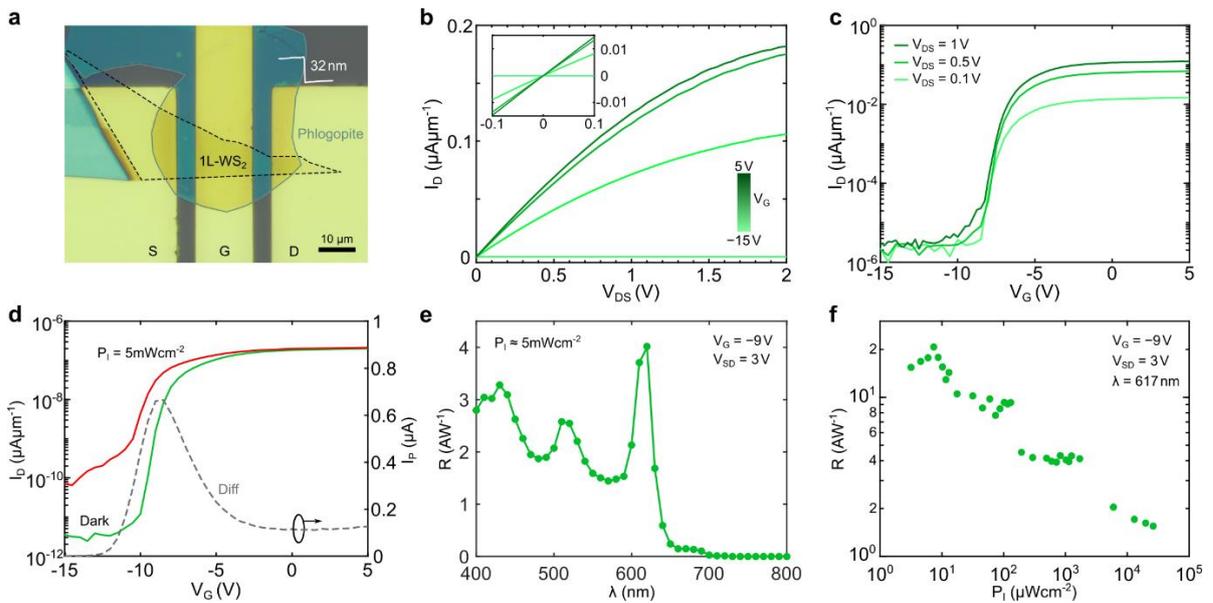

**Figure S11. Monolayer WS₂ phototransistor on phlogopite dielectric.** (a) Optical micrograph of the WS2 FET, including an AFM thickness determination. The phlogopite has a thickness of 32 nm. (b) Gate-dependent source-drain current-voltage (IVs) curves. (c) FET forward transfer characteristics for different bias voltages. (d) Pre- and during-illumination transfer characteristics of a 1L-WS₂ phototransistor under $V_{SD}$ = 1 V after exposure with a 617 nm LED with $P_i$ = 5 mWcm$^{-2}$. The panel also plots the difference between the two curves (dashed grey line), identifying the gate voltage with highest photo response ($V_G$ = –9 V). (e) Wavelength-dependent photoresponsivity under $V_G$ = –9 V and $V_{SD}$ = 3 V. (f) Illumination power-dependent photoresponsivity for a fixed gate ($V_G$ = –9 V) and source-drain bias ($V_{SD}$ = 3 V), with a maximum responsivity of 22 AW$^{-1}$.